\documentclass{article}
\usepackage{spconf,amsmath,graphicx,cite,hyperref}
\usepackage[table]{xcolor}
\usepackage{adjustbox,tabularx}
\usepackage{balance} 
\definecolor{lightgray}{rgb}{0.95,0.95,0.95}

\title{NVAlign: Direct-Gradient Optimization for Non-Verbal Control in Continuous Autoregressive Flow Matching Text-to-Speech}

\name{Qiaolin Wang$^1$, Pedro Sandoval-Segura$^1$, Anunaya Joshi$^1$, Edvardas Jurkonis$^1$, Jake Downie$^{1,\dagger}$\thanks{${}^{\dagger}$project lead.}}
\address{${}^1$Bland AI, USA}

\begin{document}\ninept
\maketitle
\raggedbottom

\begin{abstract}
While modern text-to-speech (TTS) systems generate highly natural speech and support inline non-verbal vocalization (NVV) tags, accurate control over these events remains challenging. A key gap is the lack of established post-training methods for non-verbal control in continuous autoregressive flow-matching TTS. To this end, we present \textbf{NVAlign}, a direct-gradient post-training framework for NVV tag-following in this architecture. We first perform supervised fine-tuning (SFT) of TTS models and an NVV-aware automatic speech recognition (NV-ASR) model on NVV-annotated speech, then freeze the NV-ASR model to serve as the reward model for post-training. A two-step gradient surrogate enables efficient reward backpropagation through the flow-matching sampler to jointly update the autoregressive backbone and acoustic flow head. Fidelity penalties and reference-velocity regularization help preserve speaker similarity and speech quality. Results from NVV-SuperBench and human listening evaluations show that NVAlign improves tag-following accuracy over SFT and Flow-GRPO baselines. These findings demonstrate that direct reward-gradient optimization can improve non-verbal control in continuous autoregressive flow-matching TTS. Audio samples are available at \href{https://nvalign.github.io/}{nvalign.github.io}.
\end{abstract}


\section{Introduction}
\label{sec:introduction}
\label{sec:intro}

\begin{figure*}[t]
    \centering
    \includegraphics[width=1\linewidth]{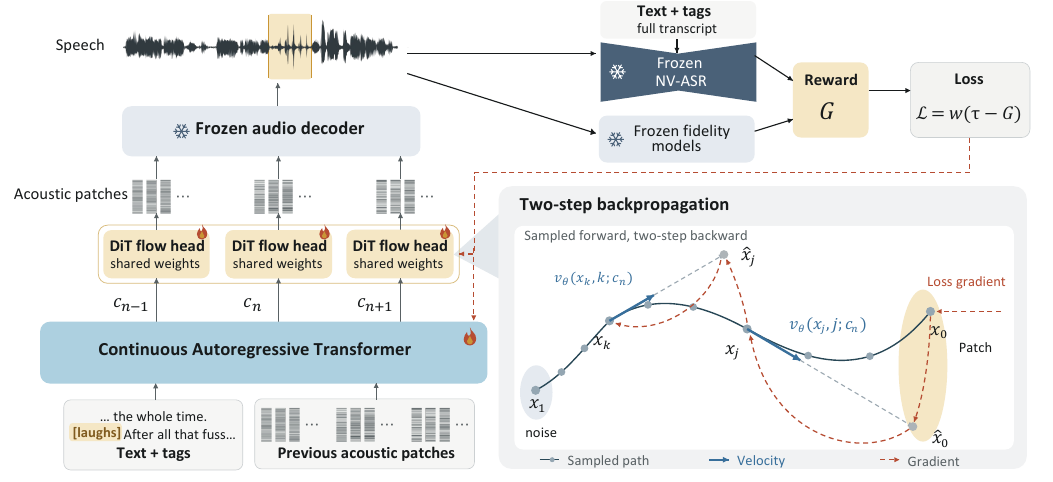}
    \caption{Overview of NVAlign post-training. Differentiable NVV rewards and fidelity penalties guide joint LoRA adaptation of the autoregressive backbone and flow head, while the decoder and evaluators remain frozen. Reference-velocity regularization (Sec.~\ref{sec:direct-gradient}) is not shown.}
    \label{fig:layout-framework}
    \vspace{-0.3cm}
\end{figure*}

A human voice conveys meaning beyond words. A laugh can soften a remark, a sigh can express reluctance, and a gasp can signal surprise. Non-verbal vocalizations shape how listeners interpret speech, making their control an important part of expressive text-to-speech. Inline NVV tags allow users to specify these events alongside spoken content \cite{elevenlabs2025audiotags,alibaba2026qwenaudiotts}, and annotated speech corpora provide supervision for their generation \cite{borisov2025nonverbaltts,liao2025nvspeech,ye2025nonverbalspeech}. However, learning the acoustic variation of vocalizations across speakers and contexts requires diverse annotated examples, which are scarce for less common sounds. Models consequently omit or confuse tagged vocalizations even after supervised fine-tuning \cite{hu2026lalmpreference,xue2026nvvsuperbench}. 

Post-training has been used to improve expressive speech and NVV generation across TTS architectures \cite{atamanenko2025tts1,liao2026fishaudios2,li2026preference}. Recent continuous autoregressive flow matching (AR-FM) TTS systems report competitive speech quality and controllable generation \cite{jia2025ditar,zhou2025voxcpm,zhou2026voxcpm2,lian2026dotstts}, motivating NVV post-training for this architecture. Unlike discrete-token TTS, where token probabilities support likelihood-ratio methods such as group relative policy optimization (GRPO) \cite{liu2025grpotts,li2026indextts25}, continuous AR-FM models use an autoregressive transformer backbone to process text tokens and preceding acoustic patches, producing continuous conditioning vectors for a diffusion transformer (DiT) flow head trained with flow matching to generate the next acoustic patch \cite{zhou2026voxcpm2,lian2026dotstts}. To accommodate continuous sampling, Flow-GRPO adapts likelihood-ratio optimization to flow models by sampling with stochastic differential equations \cite{liu2025flowgrpo,wang2026flowttsgrpo}. Existing work in continuous AR-FM TTS has also optimized reward-weighted flow losses or preference objectives \cite{yang2026grow,liu2025ardmdpo,hu2026lalmpreference}.

Continuous sampling also permits direct-gradient optimization: studies on image diffusion and flow matching models backpropagate differentiable rewards through the generation process to update model parameters \cite{clark2024draft,wu2024drtune,liang2026leapalign}. Direct-gradient optimization has also been explored across different TTS architectures, primarily targeting intelligibility, speaker identity, emotion, or representation alignment \cite{li2025dmospeech,gao2025diffro,chung2026srfd}. This raises a central question: \textbf{\textit{can an NV-ASR model provide effective reward gradients for improving tag-following in continuous autoregressive flow matching TTS?}}

To answer this question, we introduce \textbf{NVAlign}, a direct-gradient post-training framework for NVV tag-following in continuous AR-FM TTS. We first fine-tune TTS models and an NV-ASR model on NVV-annotated speech, then freeze the NV-ASR model and use its target-tag probabilities on generated speech as differentiable rewards. Using a two-step gradient surrogate~\cite{liang2026leapalign}, NVAlign efficiently backpropagates reward gradients through the flow-matching sampler to jointly adapt the backbone and flow head. Fidelity penalties and reference-velocity regularization~\cite{fang2026flowopd} help preserve speech quality and speaker similarity during this optimization. Our main contributions are as follows:
\begin{itemize}
    \item We propose NVAlign, a direct-gradient post-training framework for NVV tag-following in continuous AR-FM TTS that jointly adapts the autoregressive backbone and flow head.
    \item We formulate a differentiable NVV recognition objective with fidelity penalties and reference-velocity regularization to improve tag-following and preserve speech quality.
    \item We demonstrate gains in NVV tag-following over SFT and Flow-GRPO on NVV-SuperBench and in human listening evaluations.
\end{itemize}

\section{Related work}
\label{sec:relatedwork}

\subsection{Non-verbal vocalizations.}
Annotated NVV speech datasets support supervised training for NVV recognition and controllable speech generation~\cite{borisov2025nonverbaltts,ye2025nonverbalspeech,liao2025nvspeech,mai2025mnv17,bai2026synparaspeech,wu2025smiipnv,kanda2024elate}, while dedicated benchmarks assess recognition, generation, and perceptual quality~\cite{ni2026nvbench,xue2026nvvsuperbench,yang2026wesr,mai2026nvmos,manakul2026audiojudge}. For NVV post-training, Hu et al.~\cite{hu2026lalmpreference} use audio-language model rankings for rejection-sampling fine-tuning and Anchored Flow-DPO on VoxCPM2. NVAlign shares this focus on NVV control, but uses a frozen NV-ASR model's target-tag probabilities as differentiable rewards, backpropagating them through speech generation rather than training on utterances selected or paired by an audio-language evaluator.

\subsection{Reward and preference optimization.}
Post-training methods for diffusion and flow-based TTS differ in how reward feedback enters the optimization. GROW trains the backbone and acoustic head with reward-weighted flow matching~\cite{yang2026grow}, whereas VGPO uses gradients from a learned causal value model to update the generator~\cite{liu2026vgpo}. FlowTTS-GRPO applies likelihood-ratio updates to stochastic transitions in flow-matching acoustic models~\cite{wang2026flowttsgrpo}, while ARDM-DPO fine-tunes DiTAR using preferred and rejected speech pairs~\cite{liu2025ardmdpo}.

Direct-gradient methods backpropagate differentiable objectives through generation, as explored in image models~\cite{clark2024draft,prabhudesai2023alignprop,wu2024drtune}, with LeapAlign reducing the backward trajectory to a two-step gradient surrogate~\cite{liang2026leapalign}. In speech, DMOSpeech differentiates automatic speech recognition (ASR) and speaker objectives through a distilled diffusion model, while DiffRO uses differentiable codec-token relaxations~\cite{li2025dmospeech,gao2025diffro}. SR-FD is a close architectural precedent, optimizing a distributional speech-representation objective through VoxCPM2 and updating language-model and DiT adapters~\cite{chung2026srfd}. NVAlign extends direct-gradient speech optimization to NVV tag-following in continuous AR-FM TTS.

\section{Method}
NVAlign uses differentiable NVV recognition scores to post-train continuous AR-FM TTS (Fig.~\ref{fig:layout-framework}). Starting from supervised models, we freeze the NV-ASR model and optimize the TTS backbone and flow head through generated speech, with fidelity penalties and velocity regularization constraining the update.

\label{sec:method}

\subsection{Supervised initialization}
Since post-training primarily reinforces and recombines capabilities of the base model \cite{yuan2026f}, we first fine-tune the NV-ASR model and TTS model, parameterized by $\phi$ and $\theta$, on an NVV-annotated dataset $\mathcal D$ (Sec.~\ref{sec:eval-setup}). Each recording $a$ is paired with an $L$-token transcript $y_{1:L}$ containing words and tags such as [laughs]. The NV-ASR model $p_\phi$, based on Qwen3-Omni-30B~\cite{xu2025qwen3omni}, learns the complete transcript:
\begin{equation}
\mathcal L_{\mathrm{ASR}}(\phi)
=-\mathrm E_{(a,y)\sim\mathcal D}
\sum_{\ell=1}^{L}\log p_\phi(y_\ell\mid y_{<\ell},a).
\label{eq:asr-sft}
\end{equation}
For continuous AR-FM TTS~\cite{jia2025ditar,zhou2026voxcpm2,lian2026dotstts}, each training waveform is encoded into $N$ patches of consecutive latent frames. The continuous autoregressive transformer backbone maps tagged text, the speaker prompt, and preceding patches to a conditioning vector $c_n$ for patch $n$. A flow head $v_\theta$, shared across patches, predicts the velocity at a noised patch state $x_{n,t}$ and flow time $t$. The frozen audio decoder converts the generated patch sequence into a speech waveform $a$. The training target $u_{n,t}$ is the time derivative of the noising path:
\begin{equation}
\begin{aligned}
\mathcal L_{\mathrm{FM}}&=\mathrm E
 \big\|v_\theta(x_{n,t},t;c_n)-u_{n,t}\big\|_{\mathrm{mse}}^2,\\
\mathcal L_{\mathrm{TTS}}&=\mathcal L_{\mathrm{FM}}+\mathcal L_{\mathrm{aux}}.
\end{aligned}
\label{eq:tts-sft}
\end{equation}
The expectation averages over recordings, patches, flow times, and noise; $\|\cdot\|_{\mathrm{mse}}^2$ averages squared latent entries. We retain each TTS model's auxiliary losses, including stop-prediction losses~\cite{zhou2026voxcpm2,lian2026dotstts}.

\subsection{NVV reward and fidelity penalties}
\label{sec:method-reward}
The frozen NV-ASR model evaluates a generated waveform $a$ against its target transcript $y$ under teacher forcing: each token is scored using the audio and preceding target tokens. Let $K(y)$ be the number of NVV tags in transcript $y$, and let $S_r$ denote the token positions belonging to the $r$-th tag. We define the NVV reward as
\begin{equation}
R_{\mathrm{NVV}}(a,y)=\sum_{r=1}^{K(y)}
 \min_{\ell\in S_r}p_\phi(y_\ell\mid y_{<\ell},a).
\label{eq:nvv-reward}
\end{equation}
For a tag spanning multiple tokens, the minimum requires all of its tokens to receive high probability; summing over tags supports multiple vocalizations within an utterance.

Fidelity penalties help preserve speaker identity and audio quality during NVV reward optimization. A frozen ECAPA-TDNN~\cite{desplanques2020ecapa} extracts a speaker embedding from generated speech. Its cosine similarity to the speaker's mean embedding from SFT generations measures speaker consistency; similarity to the speaker-prompt embedding measures speaker similarity. A frozen UTMOS model~\cite{saeki2022utmos} predicts the mean opinion score (MOS) of generated speech. We penalize these three scores $q_m(a)$ below their respective bounds $b_m$. We also include signal terms for excessive waveform peaks and low signal levels:
\begin{equation}
\begin{aligned}
P_{\mathrm{fid}}(a)={}&\sum_m\lambda_m[b_m-q_m(a)]_++\lambda_{\mathrm{peak}}\sum_h[|a_h|-0.80]_+^2
\\
&+\lambda_{\mathrm{level}}\min\!\left\{[d_{\mathrm{SFT}}(y)-3-d(a)]_+,\,6\right\}.
\end{aligned}
\label{eq:fidelity-penalty}
\end{equation}
Here, $[u]_+=\max(u,0)$, all weights are nonnegative, and $a_h$ denotes the waveform amplitude at time index $h$. In dB, $d(a)$ denotes the waveform's root-mean-square (RMS) level and $d_{\mathrm{SFT}}(y)$ the median SFT level for the same text.

\subsection{Direct-gradient post-training}
\label{sec:direct-gradient}
We use the NV-ASR reward and fidelity penalties as differentiable objectives on the generated waveform, allowing their gradients to propagate through the continuous AR-FM TTS model. For each acoustic patch $n$, let $x_t=x_{n,t}$ denote a sampled state at flow time $t$ along the trajectory from noise $x_{n,1}$ to the generated patch $x_{n,0}$, conditioned on $c_n$. During post-training, we generate the patch sequence without gradients, then recompute $c_n$ using the previous acoustic patches and apply the two-step gradient surrogate~\cite{liang2026leapalign} illustrated in Fig.~\ref{fig:layout-framework}, replacing backpropagation through the full trajectory with two DiT flow-head evaluations.\par We choose flow times $1\geq k>j>0$ from the sampling grid, using the same $k$ and $j$ across patches. The first flow-head evaluation, at time $k$, estimates the state at $j$. We connect the estimate to the sampled state using a straight-through latent connector with stop-gradient $\operatorname{sg}$:
\begin{equation}
\begin{aligned}
\hat x_j&=x_k-(k-j)v_\theta(x_k,k;c_n),\\
x_j&\leftarrow\hat x_j+\operatorname{sg}(x_j-\hat x_j).
\end{aligned}
\label{eq:two-leap}
\end{equation}
The connector keeps the forward value equal to the sampled $x_j$ while routing gradients through $\hat x_j$. The second flow-head evaluation then uses $x_j$ to produce an estimate $\hat x_0$ of the generated patch $x_0$:
\begin{equation}
\begin{aligned}
\hat x_0&=x_j-jv_\theta(x_j,j;c_n),\\
x_0&\leftarrow\hat x_0+\operatorname{sg}(x_0-\hat x_0).
\end{aligned}
\label{eq:sample-connection}
\end{equation}
The frozen audio decoder maps the generated patch sequence to waveform $a$, on which the NV-ASR reward and fidelity penalties are evaluated. Reward gradients pass through the frozen evaluators and decoder to the two flow-head evaluations and, through $c_n$, to the AR backbone.

Following Flow-OPD~\cite{fang2026flowopd}, we further anchor the flow field to the SFT model by penalizing deviations from its reference velocity at the sampled states $x_k$ and $x_j$:
\begin{equation}
P_{\mathrm{vel}}=\frac{\lambda_v}{N}\sum_{n=1}^{N}\sum_{s\in\{k,j\}}
\big\|v_{n,s}-\operatorname{sg}(v^{\mathrm{ref}}_{n,s})\big\|_{\mathrm{mse}}^2.
\label{eq:velocity-anchor}
\end{equation}
Here $\lambda_v\geq0$ is the regularization weight and $v_{n,s}=v_\theta(x_{n,s},s;c_n)$ denotes the current velocity, while $v^{\mathrm{ref}}_{n,s}$ is evaluated with the SFT flow head at the same state and conditioning $c_n$ from the adapted backbone.

For a batch of $B$ utterances, we combine the NVV reward and penalties in $G_i$ and minimize:
\begin{equation}
\begin{aligned}
G_i&=R_{\mathrm{NVV}}(a_i,y_i)-P_{\mathrm{fid},i}-P_{\mathrm{vel},i},\\
\mathcal L_{\mathrm{NVAlign}}&=\frac1B\sum_{i=1}^{B}
 w_i\,(\tau_i-G_i).
\end{aligned}
\label{eq:nvalign-loss}
\end{equation}
Here, $\tau_i=K(y_i)$ is the maximum NVV reward, and $w_i$ are the detached LeapAlign weights~\cite[Eqs.~(12)--(13)]{liang2026leapalign}. We optimize low-rank adaptation (LoRA) adapters~\cite{hu2022lora} in the AR backbone, the projections mapping inputs into the flow head's feature space, and the flow head.
\begin{table*}[t]
\centering
\newsavebox{\NVMainMeasure}
\newsavebox{\NVMainTable}
\newcommand{\NVMainRows}{
\hline
\textbf{Model} & \textbf{Method} & \multicolumn{3}{c|}{\textbf{NVV-SuperBench}} & \multicolumn{2}{c|}{\textbf{Human listening sets}} & \multicolumn{3}{c}{\textbf{Quality metrics}} \\
\hline
& & \textbf{Gemini} & \textbf{Human} & \textbf{NVV detection} & \textbf{Acc.} & \textbf{MOS} & \textbf{SIM}\rlap{$^{*}$} & \textbf{UTMOS}$^{*}$ & \textbf{DNSMOS} \\
\hline
\rowcolor{white}\multicolumn{10}{l}{\textit{\textbf{Mandarin}}} \\
\rowcolor{white}VoxCPM2 & SFT & 3.36 / 2.93 & 27.4 & 26.9 / 30.2 & 43.2 & 3.11 & 0.551 & 2.44 & 2.99\\
\rowcolor{lightgray}VoxCPM2 & Flow-GRPO{\hypersetup{hidelinks}\hyperlink{fn:flow-grpo-code}{\textsuperscript{\ensuremath{\dagger}}}} & 3.04 / 2.66 & 19.2 & 27.0 / 27.7 & 37.8 & 3.25 & 0.594 & 2.82 & 3.03\\
\rowcolor{white}VoxCPM2 & \textbf{NVAlign} & 3.24 / 2.68 & \textbf{34.9} & \textbf{39.1} / \textbf{41.9} & \textbf{47.6} & 3.03 & \textbf{0.690} & \textbf{3.29} & \textbf{3.56}\\
\rowcolor{lightgray}dots.tts & SFT & 4.09 / 3.64 & 57.1 & 30.8 / 40.9 & 63.2 & 3.13 & 0.567 & 2.50 & 2.79\\
\rowcolor{white}dots.tts & Flow-GRPO{\hypersetup{hidelinks}\hyperlink{fn:flow-grpo-code}{\textsuperscript{\ensuremath{\dagger}}}} & 4.04 / 3.66 & 58.5 & 31.8 / 40.4 & 61.1 & 3.08 & 0.617 & 2.65 & 2.92\\
\rowcolor{lightgray}dots.tts & \textbf{NVAlign} & 3.76 / 2.88 & 54.4 & \textbf{49.3} / \textbf{48.5} & \textbf{65.4} & 3.04 & \textbf{0.722} & \textbf{2.97} & \textbf{3.65}\\
\hline
\rowcolor{white}\multicolumn{10}{l}{\textit{\textbf{English}}} \\
\rowcolor{white}Production TTS & SFT & 3.86 / 3.25 & 33.7 & 30.6 / 62.7 & 39.8 & 3.89 & 0.742 & 3.48 & 3.63\\
\rowcolor{lightgray}Production TTS & Flow-GRPO{\hypersetup{hidelinks}\hyperlink{fn:flow-grpo-code}{\textsuperscript{\ensuremath{\dagger}}}} & 3.84 / 3.28 & 45.3 & 29.1 / 57.0 & 31.8 & 3.92 & 0.743 & 3.56 & 3.67\\
\rowcolor{white}Production TTS & \textbf{NVAlign} & \textbf{3.95} / 2.75 & \textbf{53.7} & \textbf{42.7} / \textbf{79.9} & \textbf{54.0} & \textbf{3.95} & \textbf{0.815} & \textbf{4.06} & \textbf{3.98}\\
\hline
}
\setlength{\tabcolsep}{0pt}
\sbox{\NVMainMeasure}{\begin{tabular}{llccc|*{2}{>{\centering\arraybackslash}p{38pt}}|ccc}\NVMainRows\end{tabular}}
\setlength{\tabcolsep}{2.5pt}
\sbox{\NVMainTable}{\begin{tabular}{llccc|*{2}{>{\centering\arraybackslash}p{38pt}}|ccc}\NVMainRows\end{tabular}}
\typeout{NVALIGN-MAIN-WIDTH=\the\wd\NVMainTable; TARGET=\the\linewidth; PADDING=\the\tabcolsep}
\usebox{\NVMainTable}
\par\smallskip
\vspace{-10pt}
\caption{Results on NVV-SuperBench, human listening sets, and speech-quality metrics. Gemini reports Accuracy / PE, and NVV detection reports Whisper / SED. $^{*}$ marks metrics optimized by NVAlign.}
\label{tab:layout-wide}
\vspace{-0.2cm}
\end{table*}

\begin{table}[t]
\centering
\fontsize{9}{10.8}\selectfont
\newsavebox{\NVAbMeasure}
\newsavebox{\NVAbTable}
\newcommand{\NVAbRows}{
\hline
\textbf{VoxCPM2 variant} & \textbf{NVV reward} & \textbf{Gemini} & \textbf{SIM} & \textbf{UTMOS} \\
\hline
\rowcolor{white}SFT & 0.397 & 3.09 / 3.06 & 0.683 & 2.71\\
\hline
\rowcolor{lightgray}\textbf{NVAlign} & 0.468 & \textbf{3.54} / \textbf{3.34} & 0.798 & 3.33\\
\rowcolor{white}w/o velocity reg. & 0.488 & 2.88 / 1.94 & 0.799 & 3.39\\
\rowcolor{lightgray}w/o UTMOS & 0.474 & 2.94 / 2.28 & 0.786 & 3.01\\
\rowcolor{white}w/o signal terms & 0.454 & 2.42 / 1.52 & 0.796 & 2.72\\
\rowcolor{lightgray}w/o speaker terms & 0.457 & 2.40 / 1.31 & 0.360 & 2.31\\
\hline
\rowcolor{white}Flow head only & 0.441 & 2.89 / 3.01 & 0.767 & 3.36\\
\rowcolor{lightgray}Flow-GRPO & 0.392 & 3.01 / 3.20 & 0.712 & 2.95\\
\hline
}
\setlength{\tabcolsep}{0pt}
\sbox{\NVAbMeasure}{\begin{tabular}{@{}lc>{\hspace{3pt}}c<{\hspace{3pt}}cc@{}}\NVAbRows\end{tabular}}
\setlength{\tabcolsep}{\dimexpr(\columnwidth-\wd\NVAbMeasure)/8\relax}
\sbox{\NVAbTable}{\begin{tabular}{@{}lc>{\hspace{3pt}}c<{\hspace{3pt}}cc@{}}\NVAbRows\end{tabular}}
\typeout{NVALIGN-ABLATION-WIDTH=\the\wd\NVAbTable; COLUMN=\the\columnwidth; TEXT=9pt; PADDING=\the\tabcolsep}
\clipbox{0pt 0pt 0pt 0pt}{\usebox{\NVAbTable}}
\par\smallskip
\vspace{-10pt}
\caption{NVAlign ablations on 100 held-out VoxCPM2 texts. The middle block removes penalties cumulatively; the final two rows compare flow-head-only NVAlign and Flow-GRPO. NVV reward is averaged per tag; Gemini reports Accuracy / PE.}
\label{tab:layout-single}
\vspace{-0.2cm}
\end{table}

\vspace{-0.2cm}
\section{Experiments}
\label{sec:experiments}

\subsection{Datasets and experimental setup}
\label{sec:eval-setup}
We aggregate 834.5 hours of speech across six public NVV datasets ~\cite{borisov2025nonverbaltts,ye2025nonverbalspeech,liao2025nvspeech,mai2025mnv17,bai2026synparaspeech,wu2025smiipnv} to fine-tune VoxCPM2, dots.tts, and a Qwen3-Omni-30B-based NV-ASR model, unifying annotations into a 39-tag NVV inventory and removing benchmark text and speaker overlap. Only 18.9 hours of the public mixture are English, so for a separate production study we collect a proprietary English NVV dataset and train a separate NV-ASR model. We evaluate a production TTS model with the same continuous AR-FM architecture and NVAlign objective, using a separate internal training configuration. Within each TTS architecture, Flow-GRPO and NVAlign are initialized from the same SFT checkpoint. For VoxCPM2 and dots.tts, both methods use the NVV reward and fidelity penalties of Sec.~\ref{sec:method-reward}. Following~\cite{wang2026flowttsgrpo}, our Flow-GRPO baseline updates only the flow head with KL regularization, whereas NVAlign jointly updates the flow head and AR backbone with reference-velocity regularization.

For VoxCPM2 and dots.tts, we use LoRA ranks 16, 32, and 64 for the AR backbone, flow-head input projections, and flow head, respectively, with $\alpha=2r$. The fidelity weights are $\lambda_{\mathrm{cons}}=5$, $\lambda_{\mathrm{sim}}=\lambda_{\mathrm{UTMOS}}=\lambda_{\mathrm{peak}}=2$, and $\lambda_{\mathrm{level}}=0.5$, with $\lambda_v=0.5$. We use $b_{\mathrm{cons}}=0.75$ and $b_{\mathrm{sim}}=0.65$. For UTMOS, we set the lower bound 0.1 below the median UTMOS score of SFT generations. For each utterance, we sample $k>j$ from a 10-step flow grid and use the same pair for all patches.

We evaluate on NVV-SuperBench~\cite{xue2026nvvsuperbench}, retaining texts whose NVV tags are supported by each system. Gemini 2.5 Pro scores NVV accuracy and perceptual effect (PE) on a 0--5 scale using the official benchmark evaluation prompt. Gemini scores are comparable only within each comparison group. To compare Gemini scores with human judgments, we cover all 21 Mandarin and 19 English benchmark tags with 147 and 95 single-tag NVV-SuperBench texts, respectively, balanced across tags. Three blinded raters evaluate the generated clips; one text whose marker raters reported missing is excluded from the Mandarin VoxCPM2 subset. Whisper-large-v3~\cite{radford2023whisper}, fine-tuned on the public NVV corpus, measures tag recovery, while BEATs/ATST sound-event detection (SED)~\cite{schmid2025effective} covers cough, laugh, breath, and cry using thresholds calibrated on real recordings. 

Separate 100-text listening sets per language evaluate individual tags and tag combinations, with the Mandarin set covering all 39 tags. Using reference audio, three blinded raters judge each target sound at its marked position; tag accuracy requires agreement from all three raters. Human accuracy and NVV detection rates are reported as percentages. These sets also provide naturalness MOS (1--5), averaged over raters and clips. 

For fidelity evaluation, SIM measures ECAPA-TDNN cosine similarity between generated and speaker-reference audio, while UTMOS and DNSMOS-Pro (DNSMOS)~\cite{cumlin2024dnsmospro} estimate speech quality. We compute these metrics over benchmark outputs and 100 held-out clips per system. We bootstrap texts to obtain paired 95\% confidence intervals for human accuracy differences; Paraformer-zh~\cite{gao2022paraformer} measures CER for Mandarin and pretrained Whisper-large-v3~\cite{radford2023whisper} measures WER for English using reference transcripts without NVV tags.

\subsection{Results and analysis}
NVAlign improves tag-following over SFT and Flow-GRPO on the separate human listening sets for all three systems (Table~\ref{tab:layout-wide}), with accuracy increasing from 43.2\% to 47.6\% for VoxCPM2, 63.2\% to 65.4\% for dots.tts, and 39.8\% to 54.0\% for the English production system. On the NVV-SuperBench human subset, VoxCPM2 gains 7.5 percentage points over SFT and 15.8 points over Flow-GRPO, while dots.tts is 2.7 points below SFT. Whisper and SED detection rates likewise increase with NVAlign across all three systems. Paired 95\% bootstrap confidence intervals exclude zero for both English gains over SFT, both Mandarin VoxCPM2 gains over Flow-GRPO, and the English 100-text gain over Flow-GRPO; confidence intervals for all other comparisons include zero.

\begingroup
\hypersetup{hidelinks}
\renewcommand{\thefootnote}{\fnsymbol{footnote}}
\footnotetext[2]{\hypertarget{fn:flow-grpo-code}{}\url{https://github.com/yifan123/flow_grpo}}
\endgroup

Table~\ref{tab:layout-single} examines the contributions of joint adaptation and the penalty terms. When NVAlign is restricted to the flow head, its NVV reward still reaches 0.441, compared with 0.392 for Flow-GRPO and 0.397 for SFT; allowing joint backbone and flow-head adaptation further increases the reward and Gemini ratings. The cumulative ablations also show why the NVV objective requires additional constraints. Removing velocity regularization raises the reward from 0.468 to 0.488 while reducing Gemini accuracy/PE from 3.54/3.34 to 2.88/1.94, so greater target-tag likelihood does not necessarily correspond to better perceptual output. Removing UTMOS next lowers its predicted quality score from 3.39 to 3.01 even as Gemini ratings partially recover, and subsequently removing the peak and level terms lowers both. Removing the speaker terms then sharply reduces SIM from 0.796 to 0.360. Together, these ablations show that the different constraints control properties that are not captured by the NV-ASR reward itself.

Across the main experiments, NVAlign also increases SIM, UTMOS, and DNSMOS for all three systems. Because SIM and UTMOS are part of the training objective, we report DNSMOS and human naturalness MOS alongside them; human naturalness MOS stays within 0.1 of SFT for all three systems. Relative to SFT, Mandarin CER rises by 1.26 and 0.13 points for VoxCPM2 and dots.tts respectively; English WER rises by 0.07 on the listening set and 1.33 on NVV-SuperBench. A similar distinction appears in the English benchmark, where human tag-following rises from 33.7\% to 53.7\% while Gemini PE decreases. Tag presence, expressive quality, and speech quality therefore need not improve together.

\vspace{-0.2cm}
\section{Discussion and limitations}
\label{sec:discussion}
NVAlign is motivated by sparse supervision for less common NVVs, but the same scarcity can limit the NV-ASR reward used for post-training. Optimizing that learned reward can exploit NVV recognition shortcuts or degrade properties such as prosody because tag likelihood does not capture the full speech distribution; fidelity penalties and reference-velocity regularization constrain this drift, but adding objectives makes gradient balance itself part of the optimization problem. The current formulation also assumes discrete NVV events, leaving continuous or overlapping vocalizations outside its scope.

\vspace{-0.2cm}
\section{Conclusion}
\label{sec:conclusion}
We introduced NVAlign, a direct-gradient post-training framework for non-verbal control in continuous AR-FM TTS. By backpropagating NV-ASR rewards through a two-step flow surrogate and constraining the update with fidelity and reference-velocity penalties, NVAlign improves human tag-following on the listening sets across VoxCPM2, dots.tts, and an English production system, while also increasing Whisper tag-recovery and SED detection rates.

\clearpage
\section{Compliance with Ethical Standards}
Human listening evaluations were conducted through \href{https://www.podonos.com/}{Podonos}. No institutional ethics approval, exemption, or waiver was obtained.

\bibliographystyle{IEEEbib}
\begingroup\fontsize{9}{10}\selectfont\let\nvalignbibliography\thebibliography\renewcommand{\thebibliography}[1]{\nvalignbibliography{#1}\setlength{\itemsep}{0pt}\setlength{\parsep}{0pt}}\bibliography{refs}\endgroup

\begin{thebibliography}{10}

\bibitem{elevenlabs2025audiotags}
{ElevenLabs},
\newblock ``Introducing {Eleven} v3 (alpha),'' 2025,
\newblock
  {\urlstyle{same}\def\UrlBreaks{\do\/\do-\do\?\do\=}\url{https://elevenlabs.io/blog/eleven-v3}},
\newblock Accessed: Sep. 23, 2026.

\bibitem{alibaba2026qwenaudiotts}
{Alibaba Cloud},
\newblock ``{Qwen-Audio-3.0-TTS}: More multilingual, easier to direct,'' 2026,
\newblock
  {\urlstyle{same}\def\UrlBreaks{\do\/\do-\do\?\do\=}\url{https://www.alibabacloud.com/blog/qwen-audio-3-0-tts-more-multilingual-easier-to-direct_603379}},
\newblock Accessed: Sep. 23, 2026.

\bibitem{borisov2025nonverbaltts}
Borisov et~al.,
\newblock ``{NonverbalTTS}: A public {English} corpus of text-aligned nonverbal
  vocalizations with emotion annotations for text-to-speech,''
\newblock in {\em Proc. SSW}, 2025, pp. 104--109.

\bibitem{liao2025nvspeech}
Liao et~al.,
\newblock ``{Emilia-NV}: A non-verbal speech dataset with word-level annotation
  for human-like speech modeling,''
\newblock in {\em Proc. ICASSP}, 2026, pp. 17587--17591.

\bibitem{ye2025nonverbalspeech}
Ye et~al.,
\newblock ``A scalable pipeline for enabling non-verbal speech generation and
  understanding,''
\newblock {\em arXiv:2508.05385}, 2025.

\bibitem{hu2026lalmpreference}
Hu et~al.,
\newblock ``Preference optimization with {LALM} feedback for continuous
  autoregressive non-verbal vocalization generation,''
\newblock {\em arXiv:2609.11260}, 2026.

\bibitem{xue2026nvvsuperbench}
Xue et~al.,
\newblock ``{NVV-SuperBench}: Beyond words, beyond quality---benchmarking
  nonverbal vocalizations in speech generation,''
\newblock in {\em Proc. Interspeech}, 2026, pp. 7343--7352.

\bibitem{atamanenko2025tts1}
Atamanenko et~al.,
\newblock ``{TTS-1} technical report,''
\newblock {\em arXiv:2507.21138}, 2025.

\bibitem{liao2026fishaudios2}
Liao et~al.,
\newblock ``{Fish Audio} {S2} technical report,''
\newblock {\em arXiv:2603.08823}, 2026.

\bibitem{li2026preference}
Li et~al.,
\newblock ``Preference optimization for non-verbal vocalization synthesis,''
\newblock {\em arXiv:2608.24163}, 2026.

\bibitem{jia2025ditar}
Jia et~al.,
\newblock ``{DiTAR}: Diffusion transformer autoregressive modeling for speech
  generation,''
\newblock in {\em Proc. ICML}, 2025, pp. 27255--27270.

\bibitem{zhou2025voxcpm}
Zhou et~al.,
\newblock ``{VoxCPM}: Tokenizer-free {TTS} for context-aware speech generation
  and true-to-life voice cloning,''
\newblock {\em arXiv:2509.24650}, 2025.

\bibitem{zhou2026voxcpm2}
Zhou et~al.,
\newblock ``{VoxCPM2} technical report,''
\newblock {\em arXiv:2606.06928}, 2026.

\bibitem{lian2026dotstts}
Lian et~al.,
\newblock ``{dots.tts} technical report,''
\newblock {\em arXiv:2606.07080}, 2026.

\bibitem{liu2025grpotts}
Liu et~al.,
\newblock ``Group relative policy optimization for text-to-speech with large
  language models,''
\newblock in {\em Proc. ICASSP}, 2026, pp. 16132--16136.

\bibitem{li2026indextts25}
Li et~al.,
\newblock ``{IndexTTS} 2.5 technical report,''
\newblock {\em arXiv:2601.03888}, 2026.

\bibitem{liu2025flowgrpo}
Liu et~al.,
\newblock ``{Flow-GRPO}: Training flow matching models via online {RL},''
\newblock in {\em Proc. NeurIPS}, 2025, vol.~38.

\bibitem{wang2026flowttsgrpo}
Wang et~al.,
\newblock ``{FlowTTS-GRPO}: Online reinforcement learning with multi-objective
  reward optimization for flow-matching based text-to-speech,''
\newblock in {\em Proc. Interspeech}, 2026, pp. 1297--1306.

\bibitem{yang2026grow}
Yang et~al.,
\newblock ``{GROW}: Group-relative advantage-weighted on-policy reinforcement
  learning of autoregressive-diffusion text-to-speech model,''
\newblock {\em arXiv:2608.03215}, 2026.

\bibitem{liu2025ardmdpo}
Liu et~al.,
\newblock ``Direct preference optimization for speech autoregressive diffusion
  models,''
\newblock in {\em Proc. ICASSP}, 2026, pp. 18357--18361.

\bibitem{clark2024draft}
Clark et~al.,
\newblock ``Directly fine-tuning diffusion models on differentiable rewards,''
\newblock in {\em Proc. ICLR}, 2024.

\bibitem{wu2024drtune}
Wu et~al.,
\newblock ``Deep reward supervisions for tuning text-to-image diffusion
  models,''
\newblock in {\em Proc. ECCV}, 2024.

\bibitem{liang2026leapalign}
Liang et~al.,
\newblock ``{LeapAlign}: Post-training flow matching models at any generation
  step by building two-step trajectories,''
\newblock in {\em Proc. CVPR}, 2026, pp. 23238--23248.

\bibitem{li2025dmospeech}
Li et~al.,
\newblock ``{DMOS}peech: Direct metric optimization via distilled diffusion
  model in zero-shot speech synthesis,''
\newblock in {\em Proc. ICML}, 2025, pp. 35186--35208.

\bibitem{gao2025diffro}
Gao et~al.,
\newblock ``Differentiable reward optimization for {LLM} based {TTS} system,''
\newblock in {\em Proc. Interspeech}, 2025, pp. 2450--2454.

\bibitem{chung2026srfd}
Chung et~al.,
\newblock ``Fr{\'e}chet distance loss on speech representations for
  text-to-speech synthesis,''
\newblock {\em arXiv:2607.06027}, 2026.

\bibitem{fang2026flowopd}
Fang et~al.,
\newblock ``{Flow-OPD}: On-policy distillation for flow matching models,''
\newblock {\em arXiv:2605.08063}, 2026.

\bibitem{mai2025mnv17}
Mai et~al.,
\newblock ``{MNV-17}: A high-quality performative {Mandarin} dataset for
  nonverbal vocalization recognition in speech,''
\newblock in {\em Proc. ICASSP}, 2026, pp. 18312--18316.

\bibitem{bai2026synparaspeech}
Bai et~al.,
\newblock ``{SynParaSpeech}: Automated synthesis of paralinguistic datasets for
  speech generation and understanding,''
\newblock in {\em Proc. ICASSP}, 2026, pp. 15527--15531.

\bibitem{wu2025smiipnv}
Wu et~al.,
\newblock ``{SMIIP-NV}: A multi-annotation non-verbal expressive speech corpus
  in {Mandarin} for {LLM}-based speech synthesis,''
\newblock in {\em Proc. ACM Multimedia}, 2025, pp. 12564--12570.

\bibitem{kanda2024elate}
Kanda et~al.,
\newblock ``Making flow-matching-based zero-shot text-to-speech laugh as you
  like,''
\newblock {\em arXiv:2402.07383}, 2024.

\bibitem{ni2026nvbench}
Ni et~al.,
\newblock ``{NV-Bench}: Benchmark of nonverbal vocalization synthesis for
  expressive text-to-speech generation,''
\newblock in {\em Proc. Interspeech}, 2026, pp. 2515--2519.

\bibitem{yang2026wesr}
Yang et~al.,
\newblock ``{WESR}: A benchmark and strong baseline for word-level event-speech
  recognition,''
\newblock in {\em Findings of ACL}, 2026, pp. 3121--3134.

\bibitem{mai2026nvmos}
Mai et~al.,
\newblock ``{NVMOS}: Non-verbal vocalization quality assessment in speech,''
\newblock {\em arXiv:2606.15888}, 2026.

\bibitem{manakul2026audiojudge}
Manakul et~al.,
\newblock ``{AudioJudge}: Understanding what works in large audio model based
  speech evaluation,''
\newblock in {\em Proc. EACL}, 2026, pp. 3644--3663.

\bibitem{liu2026vgpo}
Liu et~al.,
\newblock ``{VGPO}: Fine-tuning speech autoregressive diffusion models with
  value guided policy optimization,'' 2025,
\newblock \href{https://openreview.net/forum?id=LLWIaUZvEu}{OpenReview
  preprint}.

\bibitem{prabhudesai2023alignprop}
Prabhudesai et~al.,
\newblock ``Aligning text-to-image diffusion models with reward
  backpropagation,''
\newblock {\em arXiv:2310.03739}, 2023.

\bibitem{yuan2026f}
Yuan et~al.,
\newblock ``From {$f(x)$} and {$g(x)$} to {$f(g(x))$}: {LLMs} learn new skills
  in {RL} by composing old ones,''
\newblock in {\em Proc. ICLR}, 2026, pp. 147547--147574.

\bibitem{xu2025qwen3omni}
Xu et~al.,
\newblock ``{Qwen3-Omni} technical report,''
\newblock {\em arXiv:2509.17765}, 2025.

\bibitem{desplanques2020ecapa}
Desplanques et~al.,
\newblock ``{ECAPA-TDNN}: Emphasized channel attention, propagation and
  aggregation in {TDNN} based speaker verification,''
\newblock in {\em Proc. Interspeech}, 2020, pp. 3830--3834.

\bibitem{saeki2022utmos}
Saeki et~al.,
\newblock ``{UTMOS}: {UTokyo-SaruLab} system for {VoiceMOS} challenge 2022,''
\newblock in {\em Proc. Interspeech}, 2022, pp. 4521--4525.

\bibitem{hu2022lora}
Hu et~al.,
\newblock ``{LoRA}: Low-rank adaptation of large language models,''
\newblock in {\em Proc. ICLR}, 2022.

\bibitem{radford2023whisper}
Radford et~al.,
\newblock ``Robust speech recognition via large-scale weak supervision,''
\newblock in {\em Proc. ICML}, 2023, pp. 28492--28518.

\bibitem{schmid2025effective}
Schmid et~al.,
\newblock ``Effective pre-training of audio transformers for sound event
  detection,''
\newblock in {\em Proc. ICASSP}, 2025, pp. 1--5.

\bibitem{cumlin2024dnsmospro}
Cumlin et~al.,
\newblock ``{DNSMOS Pro}: A reduced-size {DNN} for probabilistic {MOS} of
  speech,''
\newblock in {\em Proc. Interspeech}, 2024, pp. 4818--4822.

\bibitem{gao2022paraformer}
Gao et~al.,
\newblock ``{Paraformer}: Fast and accurate parallel transformer for
  non-autoregressive end-to-end speech recognition,''
\newblock in {\em Proc. Interspeech}, 2022, pp. 2063--2067.

\end{thebibliography}

\end{document}